# Low-temperature aqueous solution growth of the acousto-optic TeO$_2$ single crystals

Lu Han, Chao Liu, Xiaoli Wang, Feiyu Li, Chuanyan Fan, and Junjie Zhang*

State Key Laboratory of Crystal Materials & Institute of Crystal Materials, Shandong University, Jinan 250100, Shandong, China

E-mail: junjie@sdu.edu.cn (J. Zhang)

**Abstract:** $\alpha$-TeO$_2$ is widely used in acousto-optic devices due to its excellent physical properties. Conventionally, $\alpha$-TeO$_2$ single crystals were grown using melt methods. Here, we report for the first time the growth of $\alpha$-TeO$_2$ single crystals using the aqueous solution method below 100 °C. Solubility curve of $\alpha$-TeO$_2$ was measured, and then single crystals with dimensions of 3.5×3.5×2.5 mm$^3$ were successfully grown using seed crystals that were synthesized from spontaneous nucleation. The as-grown single crystals belong to the P4$_1$2$_1$2 space group, evidenced by single crystal X-ray diffraction and Rietveld refinement on powder diffraction. Rocking curve measurements show that the as-grown crystals exhibit high crystallinity with a full-width at half maxima (FWHM) of 57.2". Ultraviolet-Visible absorption spectroscopy indicates the absorption edge is 350 nm and the band gap is estimated to be 3.58 eV. The density and Vickers hardness of as-grown single crystals are measured to be 6.042 g/cm$^3$ and 404 kg/mm$^2$, repectively. Our findings provide an easy-to-access and energy-saving method for growing single crystals of inorganic compounds.

## 1. Introduction

Acousto-optic materials are of great interest due to their important applications in rotators, modulators, resonators, tuned filters and other acousto-optic (AO) devices.[1-7] As an excellent AO crystal, tellurium dioxide (TeO$_2$) has a melting point of 733 °C, a density of 6.0 g/cm$^3$, a large refractive index[8-11] ($n_e$=2.430, $n_o$=2.274 at a wavelength of 500 nm), and a high transmittance of visible light (transmittance of more than 70% at a wavelength of 632.8 nm, and more than 90% after coating).[12-15] More importantly, due to its large photoelastic coefficient, the speed of sound of shear waves propagating in the <110> direction is very low (only 616 m/s), and this leads to a high AO figure of merit ($M_2=n^6p^2/\rho v^{-3}$=7.93 ×10$^{-16}$ s$^3$/g), which is superior to other common AO materials such as PbMoO$_4$.[14] In addition to excellent AO properties,[16] TeO$_2$ crystals also show double-$\beta$ decay properties due to the high natural abundance of $^{130}$Te,[17, 18] which are utilized for neutrino detection.

TeO$_2$ has three polymorphs:[19-22] $\alpha$-TeO$_2$ with tetragonal rutile structure (space group $P4_12_12$),[23, 24] $\beta$-TeO$_2$ with orthorhombic slate-titanite structure (space group $Pbca$)[25] and $\gamma$-TeO$_2$ with tetragonal deformed rutile structure (space group $P2_12_12_1$).[21, 26] Among them, $\beta$-TeO$_2$ is a promising $p$-type semiconductor,[27] while $\alpha$-TeO$_2$ exhibits excellent AO properties and has attracted wide attention.[28, 29] Conventionally, melt methods such as Czochralski and Bridgman are the main techniques for single crystal growth of $\alpha$-TeO$_2$ due to congruent melting.[12, 30] In 1969, Liebertz[31] successfully grew $\alpha$-TeO$_2$ single crystals for the first time using the Czochralski method.[13, 32] $\alpha$-TeO$_2$ single crystals were mainly grown using the Czochralski method before 2006, but the $\alpha$-TeO$_2$ crystals were easy to crack, and further increase in diameter was challenging during the crystal pulling process. Later, α-TeO₂ crystals with dimensions of 52×52×80 mm³ were successfully grown using an improved Bridgman method.[12] However, both Czochralski method and Bridgman method are high-temperature crystal growth methods, which usually result in cracking issues.

In this contribution, we have successfully grown α-TeO₂ single crystals with dimensions of 3.5×3.5×2.5 mm³ for the first time using a low-temperature aqueous solution growth method. Crystal structure was determined using single crystal X-ray diffraction and confirmed by Rietveld refinement on powder X-ray diffraction pattern. Crystal quality was evaluated by Rocking curve measurements on as-grown single crystals. Physical properties including density, hardness, absorption edge and bandgap, Raman spectroscopy are characterized. Our results provide a new method for growing high quality single crystals of α-TeO₂ and other inorganic compounds with important physical properties and potential applications.

## 2. Experimental section

**2.1 Growth apparatus.** Crystal growth was performed on two different setups. For spontaneous nucleation, an oven that can be programmed to control the temperature was used. For crystal growth using seeds, the growth apparatus is shown in Figure 1a. The shell of the oil bath is made of stainless steel and has a transparent window for monitoring crystal growth. The internal container for crystallisation is made of heat-resistant glass with an inner diameter of 30 cm and a height of 50 cm, equipped with an internal stirring device with adjustable rotational speed to make the internal temperature of the solution uniform, and a temperature sensor inserted to monitor temperature. The temperature control of the growth process is realized using a Li Guan controller with an accuracy of about 0.1 °C.

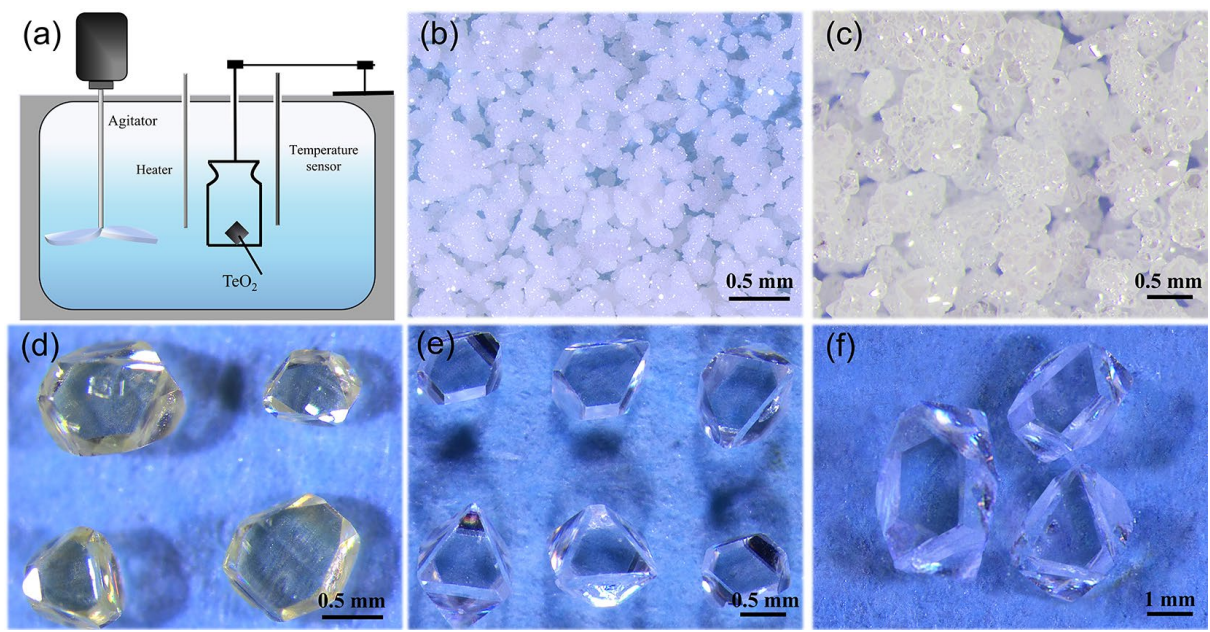

**Figure 1.** Growth apparatus and as-grown single crystals of α-TeO₂. (a) Scheme of crystal growth setup for seeded growth. (b-e) Crystals obtained from spontaneous nucleation; (f) Crystals grown using seeds.

**2.2 Crystal growth-Spontaneuous nucleation.** Spontaneous nucleation yielded transparent single crystals at the bottom of the vessel. α-TeO₂ (Aladdin, 99.999%) and CH₄N₂O (Aladdin, 99.999%) were dissolved in 150 ml of nitric acid solution (a concentration of 10%, mass ratio) in a specific ratio.

The growth vessel was sealed using cling film to minimize volatilization of solvent during crystal growth. Crystal growth was controlled using an oven with programmable temperature control. The solute was fully dissolved at 90 °C to form a clear and transparent solution, and the solution was held at this temperature for 3 hours. Crystal growth was realized by cooling the solution. Table 1 lists the growth conditions, including the ratio of each component, dwelling temperature and time, cooling range and rates, and results.

**Table 1**. Conditions and results of aqueous solution growth of $\alpha$-TeO$_2$.

| No. | $\alpha$-TeO$_2$:CH$_4$N$_2$O (Mass Ratio) | HNO$_3$ (10%wt), (ml) | Dwelling Temperature (°C) and time (h) | Cooling Range (°C), Rate (°C/h) | Result |
|---|---|---|---|---|---|
| 1 | 1.9: 20 | 150 | 90, 3 | 90-60, 0.21 | Powders (crystals ≤0.05 mm on edge), Figure 1b |
| 2 | 3: 0 | | | 90-60, 0.21 | Powders (crystals ≤0.05 mm on edge) |
| 3 | 3: 10 | | | 90-60, 1.6 | Powders (crystals ≤0.05 mm on edge), Figure 1c |
| 4 | 3: 20 | | | 90-80, 0.07 | Yellow crystals (0-1 mm on edge), Figure 1d |
| 5 | 3: 20 | | | 90-60, 0.21 | Colorless crystals (0-1 mm on edge) Figure 1e |
| 6 | 3.6:20 | | | 90-60, 0.21 | Powders (crystals ≤0.05 mm on edge) |

**2.3 Crystal growth-Seeded growth.** Figure 2 shows the saturation temperatures for different concentration. The saturation temperature was measured by observing whether a seed crystal grows or dissolves at various temperatures for a specific concentration. Specifically, a clear and transparent solution was prepared at 90 °C. The temperature is then gradually lowered by 5 °C at a time. For instance, when the temperature reaches 80 °C, and kept constant. A piece of $\alpha$-TeO$_2$ single crystal was dipped in the solution for a couple of hours. If the crystal becomes smaller, it indicates that the temperature is above the saturation temperature and needs to be lowered further. If the crystal grows, it indicates that the temperature is above the saturation point and the solution goes back to 90 °C. This process needs to repeat multiple times until we find the temperature at which the seed crystal does not grow or dissolve, and this is the saturation temperature.

The solubility curve was obtained by fitting with third polynomial function: $y = a+bT+cT^2$, where $a$=2.854g/L, $b$=2.794×10$^{-2}$g/(L·°C) and $c$=8.269×10$^{-4}$g/(L·°C$^2$). For seeded growth, $\alpha$-TeO$_2$ (Aladdin, 99.999%) and CH$_4$N$_2$O (Aladdin, 99.999%) were mixed in a mass ratio of 3:20 and dissolved in 150 ml of nitric acid solution with a concentration of 10% (mass ratio). The growth vessel was sealed. The solute was completely dissolved by heating the solution to 90 °C and keeping it for 3 hours to form clear and transparent solution. The saturation temperature was measured by observing the growth or dissolution of the seed crystal. Then, the seed crystals were placed in the solution 3-5 °C above the saturation temperature. Crystal growth occurs in the temperature range of 76-70 °C with a growth time of 7-24 days.

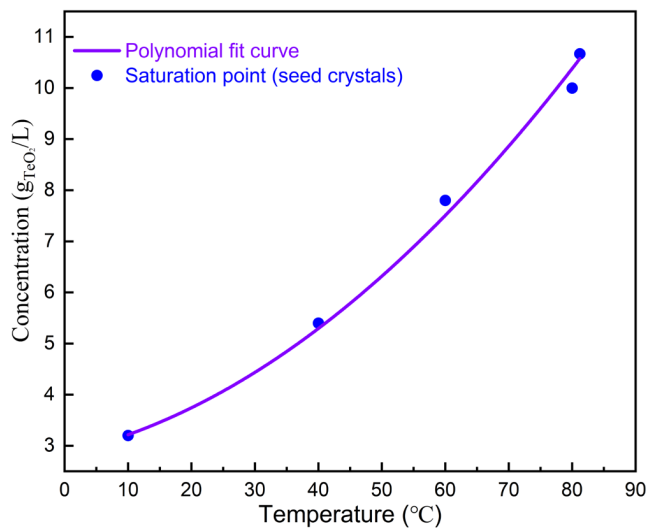

**Figure 2.** Saturation temperature as a function of concentration for $\alpha$-TeO$_2$.

**2.4 In-house X-ray powder diffraction (P-XRD).** The as-grown crystals were thoroughly ground and analyzed by X-ray powder diffraction using a Bruker AXS D2 Phaser X-ray powder diffractometer. Data were collected with an X-ray wavelength of 1.5418 Å in the 2θ range of 5-140° with a step size of 0.02° and a step time of 1.2 s. Rietveld refinement were carried out using TOPAS 6. Refined parameters include background (Chebyshev function, 5 order), zero error, lattice parameters, crystal size L, and strain G.

**2.5 Single crystal diffraction.** Single crystal X-ray diffraction data were collected on a Bruker AXS D8 Venture diffractometer at 298 K (Mo-K$_{\alpha 1}$ radiation, λ = 0.71073 Å). A piece of single crystal with dimensions of 0.045×0.049×0.050 mm$^3$ was used. Indexing was performed using Bruker APEX4 software.[33] Data integration and cell refinement were performed using SAINT, and multiscan absorption corrections were applied using the SADABS program. The structure was solved with the XT[34] structure solution program using Intrinsic Phasing and refined with the XL[35] refinement package using Least Squares minimisation. All atoms were modelled using anisotropic ADP, and the refinements converged for $I > 2\sigma(I)$, where $I$ is the reflected intensity and $\sigma(I)$ is the standard deviation. Calculations were performed using SHELXTL[34] and Olex2.[36] Further details of the crystal structure investigations may be obtained from the joint CCDC/FIZ Karlsruhe online deposition service by quoting the deposition number CSD 2295591.

**2.6 High-resolution X-ray diffraction.** To determine the quality of the as-grown crystals, rocking curve was measured. The size of the crystals used in the experiment was 3×3×2.5 mm$^3$. Data were collected at room temperature using a Cu target with a minimum step of 0.0001° on a Smart Lab 3kW model from Japan.

**2.7 Density.** The density of α-TeO$_2$ were measured using the Archimedes method at 24 °C. The formula used were $\rho_{exp} = m_0 \rho_{water}/(m_0 - m_1)$, where $m_0$ is the sample weight in air, $m_1$ is the sample weight immersed in distilled water, and $\rho_{water}$ is the density of distilled water at 24 °C ($\rho_{water}$=0.99730 g/cm$^{-3}$). The final density was obtained by calculating the average of three measurements. α-TeO$_2$ was measured to have a density of 6.042 g/cm$^3$.

**2.8 Hardness.** The mechanical hardness of the grown α-TeO$_2$ single crystals was measured using a fully automated microhardness tester (model: HZ52-4), with an indentation load of 5 g, an application time of 2 s, and a selection of 5-7 points to be averaged as the hardness data of the α-TeO$_2$ single crystals. α-TeO$_2$ has a Vickers hardness of 404 kg/mm$^2$.

**2.9 Ultraviolet-Visible Absorption Spectroscopy (UV-Vis).** In order to determine the UV cut-off edge and its band gap, we collected UV-Vis data on pulverized single crystals at room temperature using a model 8453E UV-Vis spectrometer with a resolution of >1.6 nm and a scanning range of 190-1100 nm.

**2.10 Raman Spectroscopy.** Data were collected using single crystals with dimensions of 0.4×0.4×0.5 mm$^3$ on a Raman spectrometer model PHS-3C equiped with a laser with a wavelength of 473 nm.

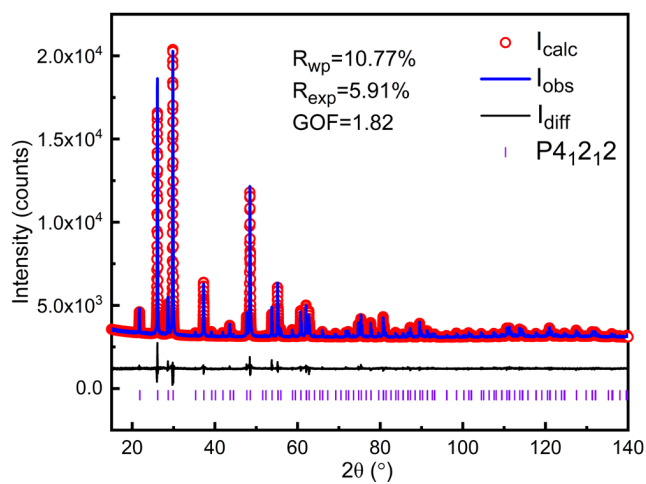

**Figure 3.** Rietveld refinement on in-house powder X-ray diffraction pattern ($\lambda$ = 1.5418 Å) of pulverized as-grown crystals of α-TeO$_2$. $I_{obs}$: observed intensity; $I_{diff}$: difference; $I_{calc}$: calculated intensity.

## 3. Results and Discussion

**3.1 Crystal growth.** α-TeO$_2$ has been commercially used in various acousto-optic devices and single crystals of this material are tranditionally grown using Czochralski and Bridgman methods due to its congruent melting and high efficiency.[12, 30] However, cracking issues exist due to stress formed from sharp temperature gradient and relatively fast growth rate. Here we grow high quality single crystals of α-TeO$_2$ using the aqueous solution method for the first time.

Initially, we explored the growth of α-TeO$_2$ using hydrothermal method; however, no α-TeO$_2$ were found. Then, we noticed a paper reporting the growth of β-Ga$_2$O$_3$ single crystals at low temperature using nitric acid as a co-solvent and urea as a nucleating agent.[37] We immediately tried this method. α-TeO$_2$ and $CH_4N_2O$ in the mass ratio of 1.9:20 was dissolved in 150 ml of nitric acid solution with a concentration of 10% (mass ratio). These materials form a homogeneous solution at 90 °C. We then cooled the solution from 90 to 60 °C using the cooling rate of 0.21 °C /h. Polycrystalline powders were obtained by removing the solution (see Table 1 and Figure 1b). Figure 3 shows the in-house powder diffraction data. All peaks match with $α$-$TeO_2$ (PDF # 04-007-2021) and can be indexed using the space group $P4_12_12$. Rietveld refinement was performed using the structural model from single crystal X-ray diffraction (discussed below) as a starting point. The refinement converged to $R_{wp}$ =10.77%, $R_p$=8.23%, $R_{exp}$=5.91% and GOF=1.82 with lattice parameters of $a=b$=4.8107(1) Å, $c$=7.6126(1) Å, which are comparable to the previous results.[38]

We turn to optimize the growth conditions based on our preliminary experimental results, including the content of each component, the growth temperature, and the degree of sealing (see Table 1) to improve the size and quality of single crystals. We first tuned the ratio of $α$-$TeO_2$, $CH_4N_2O$ and $HNO_3$ while keeping other conditions unchanged. The results showed that fixing the nitric acid concentration at 10% (mass ratio) had a better effect on the yield and quality of the crystals when the mass ratio of $α$-$TeO_2$ and $CH_4N_2O$ was 3:20 (see Table 1, 5). However, when the ratio of $α$-$TeO_2$ to $CH_4N_2O$ was less than 2:20 or more than 3.5:20, polycrystalline powder tends to be formed (see Figure 1b), and in addition, if the ratio of each component was appropriate, polycrystalline powder can be generated if the growth time was insufficient (see Figure 1c). Meanwhile, we found that high temperature caused excessive solvent volatilisation, making it challenging to control the crystal quality and growth rate.

In order to solve the problem of quick evaporation at high temperatures, we made some adjustments. A screw cap and safety film were used for double sealing, but three small holes with a diameter of about 500 μm were intentionaly made in the cling film. By doing so, the volatilisation rate of the solution was reduced, providing a stable environment for crystal growth. The most suitable growth conditions were achieved by adjusting the ratio of each component in the solution. Currently, excellent quality and size of crystals can be grown when the mass ratio of $α$-$TeO_2$ and $CH_4N_2O$ is around 3:20. By optimizing the cooling rate, high-quality single crystals with dimensions up to 1 mm on edge can be obtained through spontaneous crystallization (see Figure 1e). These small single crystals can be used as seed crystals for further growth.

During the experiments, it was found that the color of the grown crystals was yellowish when the growth temperature was above 80 °C (see Figure 4a). However, it was found that yellow crystals can be changed into colorless translucent crystals after annealing at 600 °C in air (see Figure 4b). First, we carried out powder XRD measurement on the as-grown yellow crystals, and find that it matches well with $α$-$TeO_2$ (PDF # 04-007-2021) and no other peaks appear. Rietveld refinements (see Figures 4c, d) on the pulverized crystals before and after annealing using the single crystal structural model show that the unit cell parameters do not change before and after annealing ($a=b$=4.8121(4) Å, $c$=7.6150(3) Å before annealing and $a=b$=4.8120(3) Å, $c$=7.6146(4) Å after annealing). There are two possibilities for the yellow colour: (1) tiny impurities below our resolution and (2) inclusions in the crystal that are amorphous.

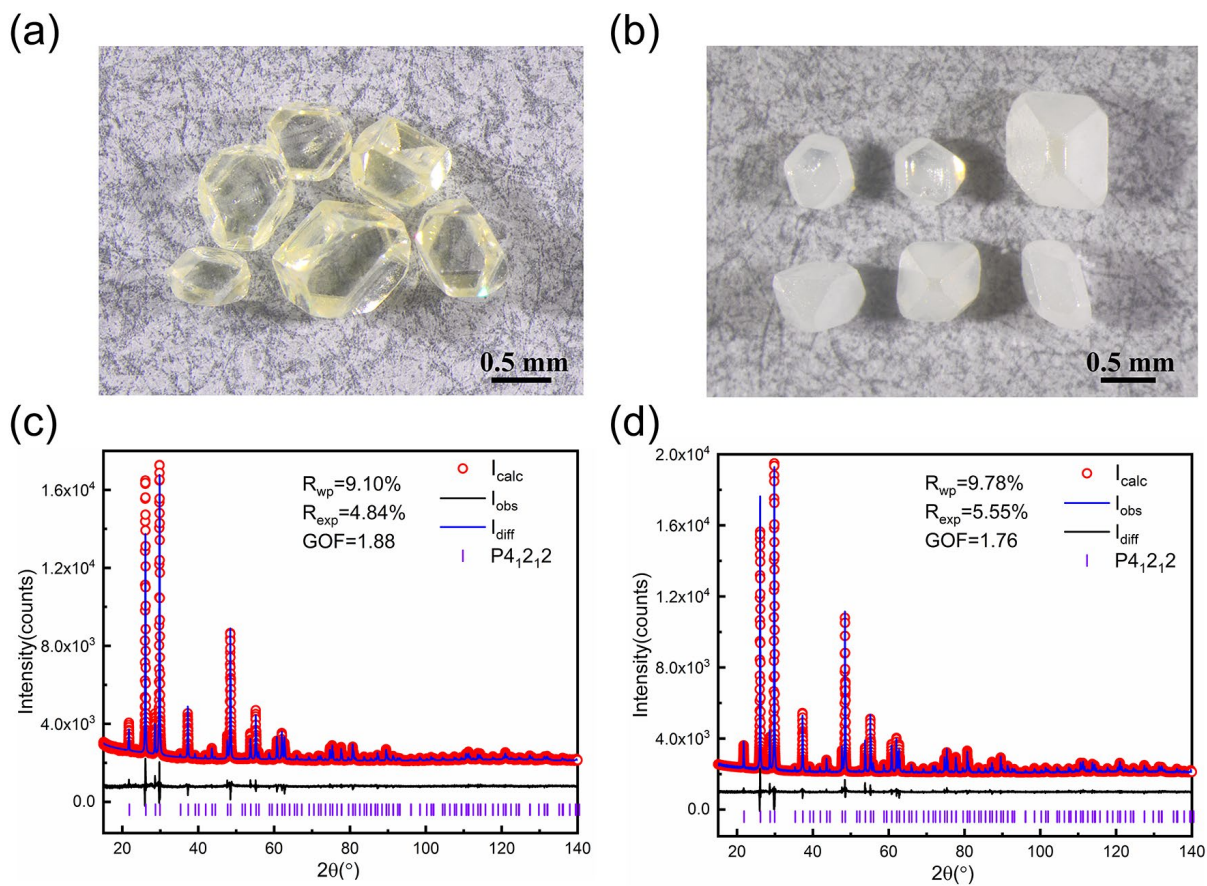

**Figure 4.** (a) As-grown light-yellow crystals, (b) Colourless semi-transparent crystals after annealing in air at 600 °C; (c) Rietveld refinement on powder X-ray diffraction pattern of pulverized as-grown light-yellow crystals; (d) Rietveld refinement on powder X-ray diffraction pattern of pulverized crystals after annealing in air at 600 °C.

We then used the single crystals obtained from spontaneous nucleation as seeds for single crystal growth. The saturation temperature as a function of concentration is shown in Figure 2. As can be seen, the saturation temperature increases as the concentration increases. We chose conditions when the $\alpha$-$TeO_2$ and $CH_4N_2O$ mass ratio was 3:20 for crystal growth (see Table 1, 5). According to the solubility curve, the saturation temperature is expected to be 74 °C, consistent with 75 °C from measurements. A seed crystal was dipped at 79 °C, the solution was cooled to 75 °C at a rate of 0.33 °C/h in order to slightly dissolve the surface of the seed crystal, and then the solution was further cooled to 70 °C at a rate of 0.01-0.017°C/h to grow crystal. After a growth period of 16 days, single crystals with dimensions of 3.5×3.5×2.5 mm$^3$ (see Figure 1f) were successfully obtained.

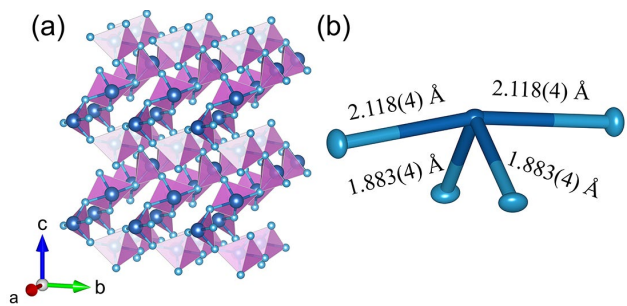

**Figure 5.** Crystal structure of α-TeO$_2$. (a) Three-dimensional structure using the polyhedral model; (b) Local environment of Te using the ellipsoid model (50% probility) with bond distances labeled.

**3.2 Single crystal diffraction.** We performed single crystal diffraction on an α-TeO$_2$ single crystal with dimensions of 0.045×0.049×0.050 mm$^3$. Figure 5a shows the three-dimensional crystal structure of α-TeO$_2$, which belongs to the tetragonal $P4_12_12$ space group. There are one Te atom and one O atom in the asymmetric unit. The Te atom is surrounded by four oxygen atoms to form a seesaw geometry due to the existence of lone pairs of electrons in Te (see Figure 5b), as previous observed in tellurites.[39] The bond distances of Te-O are in the range of 1.883(4)-2.118(4) Å. Adjacent TeO$_4$ polyhedra are connected to each other by sharing corners to form a three-dimensional network. Our structure is consistent with previous reports.[40]

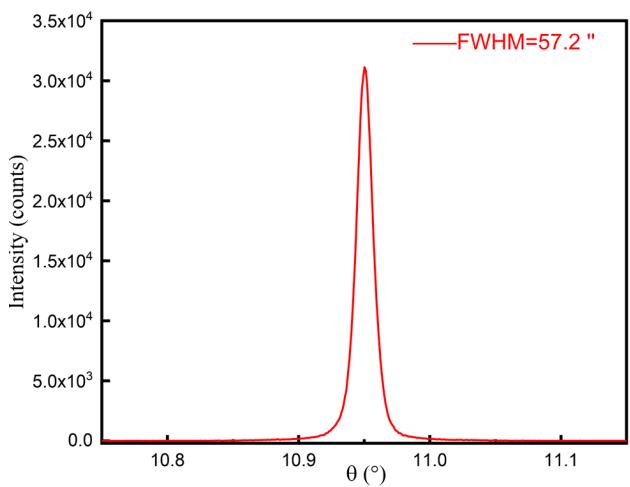

**Figure 6.** Rocking curves of the (101) plane of a typical as-grown sample.

**3.3 Rocking curve.** A piece of single crystal with (101) polished was used to measure the rocking curve. The theta-scale was scanned from 10.782 to 11.218° with scan steps of 0.001°. The curve, as shown in Figure 6, shows nice symmetry and smoothness and a small value of FWHM (57.2"), indicating high crystallinity of as-grown α-TeO$_2$ single crystals.

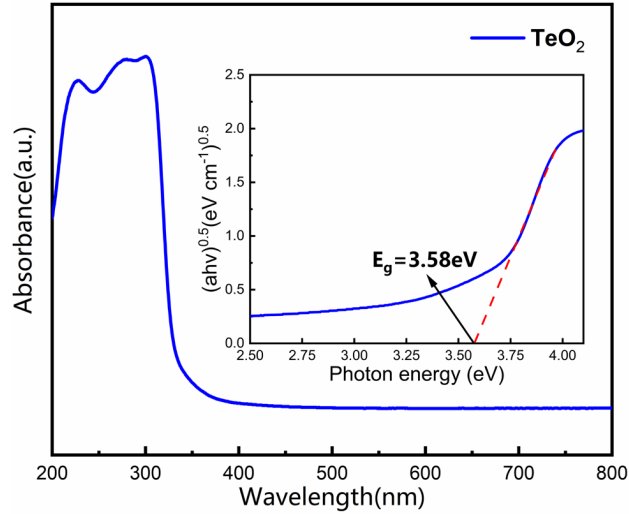

**Figure 7.** Ultraviolet-Visible absorption spectra and band gap of pulverized α-TeO$_2$ single crystals.

**3.4 Ultraviolet-Visible Absorption Spectroscopy (UV-Vis).** In order to examine the optical properties of our grown α-TeO$_2$ single crystals, we performed UV-visible spectral analysis by UV diffuse reflectance testing on the as-grown crystals (see Figure 7). It is evident that the UV cut off for visible α-TeO$_2$ crystals is around 350 nm, which is also in agreement with the data documented in the literature.[14] The absorption edge and band gap for semiconductors can be explained by a power law $(\alpha h\nu)^{0.5}=A(h\nu-E_g)$, where α, h, ν, A and $E_g$ denote the absorption coefficient, Planck's constant, incident light frequency, constant, and optical band gap, respectively. The inset in Figure 7 shows the band gap of α-TeO$_2$, which is estimated to be 3.58 eV, in good agreement with the literature.[21]

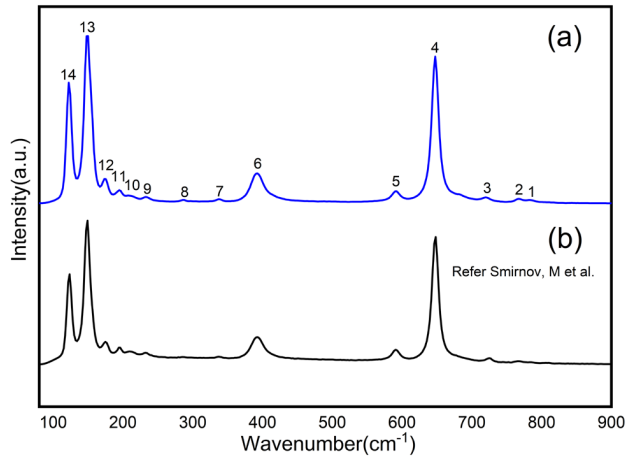

**Figure 8.** Comparison of experimental (a) and reported (b) Raman spectra of α-TeO$_2$.

**3.5 Raman spectrum.** The Raman spectrum of α-TeO$_2$ can be divided into three different groups of peaks (see Figure 8), peaks 1-5 in the high-frequency region, peaks 6-8 in the mid-frequency region and peaks 9-14 in the low-frequency region. The peaks in the high-frequency region and peaks in the mid-frequency region can be attributed to the Vas (Te-O-Te) and Vs (Te-O-Te) modes, respectively, and the peaks in the low-frequency region are mixed modes including the O-Te-O angular bending and the Te-O-Te bridge rotational oscillations. The Raman spectrum of our as-grown single crystals are consistent with the report from Smirnov,[41] demonstrating high quality of our as-grown crystals.

## 4. Conclusions

We report for the first time the successfully growth of millimeter-sized single crystals of acousto-optic $\alpha$-$TeO_2$ using spontaneous nucleation and seeded growth using the low-temperature aqueous solution growth method. The as-grown single crystals belong to the tetragonal $P4_12_12$ space group and form a three-dimensional crystal structure via coner-sharing of $TeO_4$ polyhedra. Rocking curve measurements show high crystallinity of as-grown single crystals with a FWHM of 57.2″. Physical properties including density, hardness, UV absorption edge and band gap are characterized. Our findings provide a completely new method for growing single crystals of other interesting inorganic compounds with potential in scientific research and application.

## Accession Codes

CCDC 2295591 contains the supplementary crystallographic data for this paper. These data can be obtained free of charge via www.ccdc.cam.ac.uk/data_request/cif, by emailing at data_request@ccdc.cam.ac.uk, or by contacting the Cambridge Crystallographic Data Centre, 12 Union Road, Cambridge CB2 1EZ, UK; fax: +44 1223 336033.

## AUTHOR INFORMATION

Corresponding Author

Junjie Zhang – State Key Laboratory of Crystal Materials and Institute of Crystal Materials, Shandong University, Jinan, Shandong 250100, China; orcid.org/0000-0002-5561-1330; Email: junjie@sdu.edu.cn

## Author Contributions

The manuscript was written through contributions of all authors. All authors have given approval to the final version of the manuscript.

## Notes

The authors declare no competing financial interest.

## Acknowledgements

J.Z. thanks Prof. Xutang Tao for providing valuable support and fruitful discussions. J. Z and L.H thank Prof. Jian Zhang for his help with in-house single crystal X-ray diffraction. J.Z thanks Dr. Yu-Sheng Chen and Dr. Tieyan Chang from The University of Chicago for stimulating discussions. Work at Shandong University was supported by the National Natural Science Foundation of China (12074219 and 12374457), the TaiShan Scholars Project of Shandong Province (tsqn201909031), the QiLu Young Scholars Program of Shandong University, the Crystalline Materials and Industrialization Joint Innovation Laboratory of Shandong University and Shandong Institutes of Industrial Technology (Z1250020003), and the Project for Scientific Research Innovation Team of Young Scholars in Colleges and Universities of Shandong Province (2021KJ093).# References

(1) Peercy, P. S.; Fritz, I. J. Pressure-Induced Phase Transition in Paratellurite (TeO$_2$). *Physical Review Letters* **1974**, *32* (9), 466-469.
(2) Fritz, I. J.; Peercy, P. S. Phenomenological theory of the high-pressure structural phase transition in paratellurite (TeO$_2$). *Solid State Communications* **1975**, *16* (10-11), 1197-1200.
(3) Skelton, E. F.; Feldman, J. L.; Liu, C. Y.; Spain, I. L. Study of the pressure-induced phase transition in paratellurite (TeO$_2$). *Physical Review B* **1976**, *13* (6), 2605-2613.
(4) Korablev, O.; Bertaux, J. L.; Grigoriev, A.; Dimarellis, E.; Kalinnikov, Y.; Rodin, A.; Muller, C.; Fonteyn, D. An AOTF-based spectrometer for the studies of Mars atmosphere for Mars Express ESA mission. *Advances in Space Research* **2002**, *29* (2), 143-150.
(5) Tran, C. D. Principles, Instrumentation, and Applications of Infrared Multispectral Imaging, An Overview. *Analytical Letters* **2005**, *38* (5), 735-752.
(6) Mantsevich, S. N.; Korablev, O. I.; Kalinnikov, Y. K.; Ivanov, A. Y.; Kiselev, A. V. Wide-aperture TeO$_2$ AOTF at low temperatures: Operation and survival. *Ultrasonics* **2015**, *59*, 50-58.
(7) Maák, P.; Barócsi, A.; Fehér, A.; Veress, M.; Mihajlik, G.; Rózsa, B.; Koppa, P. Acousto-optic deflector configurations optimized for multiphoton scanning microscopy. *Optics Communications* **2023**, *530*, 1-10.
(8) Uchida, N.; Ohmachi, Y. Elastic and Photoelastic Properties of TeO2 Single Crystal. *Journal of Applied Physics* **1969**, *40* (12), 4692-4695.
(9) Dafinei, I.; Diemoz, M.; Longo, E.; Péter, Á.; Földvári, I. Growth of pure and doped TeO$_2$ crystals for scintillating bolometers. *Nuclear Instruments and Methods in Physics Research Section A: Accelerators, Spectrometers, Detectors and Associated Equipment* **2005**, *554* (1-3), 195-200.
(10) Tagiara, N. S.; Palles, D.; Simandiras, E. D.; Psycharis, V.; Kyritsis, A.; Kamitsos, E. I. Synthesis, thermal and structural properties of pure TeO$_2$ glass and zinc-tellurite glasses. *Journal of Non-Crystalline Solids* **2017**, *457*, 116-125.
(11) Moufok, S.; Kadi, L.; Amrani, B.; Khodja, K. D. Electronic structure and optical properties of TeO$_2$ polymorphs. *Results in Physics* **2019**, *13*, 1-5.
(12) Chu, Y.; Li, Y.; Ge, Z.; Wu, G.; Wang, H. Growth of the high quality and large size paratellurite single crystals. *Journal of Crystal Growth* **2006**, *295* (2), 158-161.
(13) Kokh, A. E.; Shevchenko, V. S.; Vlezko, V. A.; Kokh, K. A. Growth of TeO$_2$ single crystals by the low temperature gradient Czochralski method with nonuniform heating. *Journal of Crystal Growth* **2013**, *384*, 1-4.
(14) Zeng-Wei, G. E.; Xue-Ji, Y. I. N.; Wei, W.; Shi-Hai, Y. U. E.; Yong, Z. H. U. Research Progress on TeO$_2$ Crystal and Its Utilization in Infrared Devices. *Journal of Inorganic Materials* **2015**, *30* (8), 802-808.
(15) Amari, A.; Al Mesfer, M. K.; Alsaiari, N. S.; Danish, M.; Alshahrani, A. M.; Tahoon, M. A.; Rebah, F. B. Electrochemical and Optical Properties of Tellurium Dioxide (TeO$_2$) Nanoparticles. *International Journal of Electrochemical Science* **2021**, *16* (2), 1-10.
(16) Tian, Y.; Chen, Y.; Song, D.; Liu, X.; Bi, S.; Zhou, X.; Cao, Y.; Zhang, H. Acousto-optic tunable filter-surface plasmon resonance immunosensor for fibronectin. *Analytica Chimica Acta* **2005**, *551* (1-2), 98-104.
(17) Qin, B.; Bai, Y.; Zhou, Y.; Liu, J.; Xie, X.; Zheng, W. Structure and characterization of TeO$_2$ nanoparticles prepared in acid medium. *Materials Letters* **2009**, *63* (22), 1949-1951.
11

Table of Contents only

We report for the first time the growth of high-quality single crystals of acousto-optic $\alpha$-TeO$_2$ using the aqueous solution method.

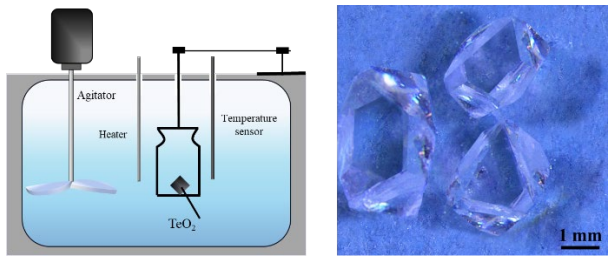